\documentclass[aps,prl,showpacs,amsmath,amssymb,floatfix,superscriptaddress,twocolumn,longbibliography]{revtex4-2}

\usepackage{times,graphicx,color,placeins}
\usepackage[colorlinks=true,urlcolor=blue,linkcolor=blue,citecolor=blue]{hyperref}

\begin{document}
\title{Quadriexcitons and excitonic condensate in a symmetric electron-hole bilayer with valley 
degeneracy}

\author{Stefania De Palo}
\altaffiliation{depalo@iom.cnr.it}
\affiliation{CNR-IOM-DEMOCRITOS, Trieste, Italy}
\affiliation{Dipartimento di Fisica,
Universit\`a di Trieste, strada Costiera 11, 34151
Trieste, Italy}

\author{F. Tramonto}    
\altaffiliation[Present address: ]{Kyndryl Italia Innovation Services S.r.l., Via Circonvallazione Idroscalo snc, 20090 Segrate (MI), Italy}
\affiliation{Dipartimento di Fisica,
Universit\`a di Trieste, strada Costiera 11, 34151
Trieste, Italy}

\author{Saverio Moroni}
\altaffiliation{moroni@iom.cnr.it}
\affiliation{CNR-IOM-DEMOCRITOS, Trieste, Italy}
\affiliation{SISSA (International School for Advanced 
Studies), via Bonomea 265, 34136 Trieste, Italy}

\author{Gaetano Senatore}
\altaffiliation{senatore@units.it}
\affiliation{Dipartimento di Fisica,
Universit\`a di Trieste, strada Costiera 11, 34151
Trieste, Italy}

\begin{abstract}
 Using quantum Monte Carlo simulations we have mapped out the zero temperature  phase 
 diagram of a symmetric electron-hole bilayer with twofold valley  degeneracy, as function of the 
 interlayer distance $d$ and in-layer density $n$. We find  that the effect of the valley degeneracy 
 is to shrink the region of stability  of the  excitonic  condensate, in favor of quadriexcitons at  
 small $d$ and of the  four-component   plasma at large  $d$, with minor effects on  the value of  
 the excitonic condensate fraction.  The enclosure of the condensate in a density window possibly 
 explains why anomalous  tunnelling  conductivity, interpreted as signature of  condensation,  
 is observed only between two  finite  values of carrier density  in graphene bilayers. 
 Our phase diagram may provide directions to select device parameters for future experiments. 
\end{abstract}

\date{\today} 

\maketitle

The idea that spatially separated electrons and holes  provide an 
ideal playground for the observation of  superfluidity/superconductivity was put forward long ago 
\cite{lozovik1975,*lozovik1976}.  An equilibrium condensate, however,   has long remained elusive in 
conventional  electron-hole semiconductor bilayers \cite{prlsivan1992,prlcroxall2008,prlseamons2009},  
while it  was observed in semiconductor  electron-electron (hole-hole)   bilayers \cite{kelloggprl2004,tutucprl2004,jimallan2004} 
 in strong perpendicular magnetic fields \cite{fert89},  i.e., in the quantum Hall regime.  It was recently argued  that  excitonic 
condensation should be strongly  enhanced  in coupled electron-hole graphene bilayers, where extremely thin hBN barriers can be used  
\cite{peralietal13} and electrons and holes have equal masses.  Yet,  carrier valley degeneracy is present in 
graphene bilayers and its effect  is not  immediately  obvious; moreover, (i)  BCS mean field \cite{prllittlewood1995} would 
suggest  an enhancement of the condensate,  (ii) the screening by a larger number of Fermion components 
would  point to a weakening of the e-h attraction, and (iii)  the presence of four Fermion components  in each bilayer would 
allow for the formation of quadriexcitons. \cite{kittel1972}. 

Experiments on coupled  graphene bilayers promptly followed the proposal in Ref. \onlinecite{peralietal13}, 
replicating however the scenario encountered in semiconductor bilayers, with  no evidence of a condensate in 
coupled electron-hole graphene bilayers \cite{deanGBprl,tutucGBprl} and a clear evidence of a  condensate in  
coupled electron-electron graphene bilayers in  the quantum Hall regime \cite{liunpCGB2017, linpCBG2017}. 
Eventually, evidence of condensed interbilayer excitons in zero magnetic field, signaled by enhanced tunneling, 
was reported  in coupled electron-hole graphene bilayers \cite{tutucprl2018} and more recently in coupled monolayers  
of transition metal dichalcogenides \cite{faimak2019, fogler}; moreover, thermodynamic evidence of the 
condensate was also provided \cite{faimak2021}. We should recall at this point that some  evidence of condensation  was also  
found in semiconductor electron bilayers,  with long lived indirect excitons produced by photoexcitation  and electrostatic  trapping \cite{butovnano2012,PRLDubin18} and in excitons coupled to light confined within a cavity (exciton-polaritons), which  
do live long enough to condense  and  require a continuous input of light \cite{polaritons}.

Here, we restrict to equilibrium excitonic condensation  in systems of electron and holes and assess the effects of 
the interplay of the valley degeneracy $g_v$,  the interlayer distance $d$ and the in-layer density $n$ in 
determining  the $T=0$  phase diagram of the system.  To this end we  resort to  QMC simulations of the simplest 
possible model, i.e., a paramagnetic, symmetric electron-hole bilayer ($m_e=m_h=m_b$,  $g_v=2$),  to mimic 
the situation encountered in double bilayer graphene. In effective Hartree atomic units ($\hbar = 
m_b=e^2/(4\pi\varepsilon_0\epsilon)=1$), which we use throughout, the Hamiltonian  of the system reads
\begin{eqnarray}
H=&-&\frac{1}{2}\sum_{v,i}{\nabla^2_{e,v,i}}  +\frac{1}{2} \sideset{}{'}\sum_{v,v',i, i'}\frac{1}{|{\bf r}_{e,v,i} -{\bf r}_{e,v',i'}|}  \nonumber\\
&-&\frac{1}{2}\sum_{v,i}{\nabla^2_{h,v,i}}+\frac{1}{2} \sideset{}{'}\sum_{v,v',i, i'}\frac{1}{|{\bf r}_{h,v,i} -{\bf r}_{h,v',i'}|}  \nonumber\\
 &-&\sum_{v,v',i, i'}\frac{1} {\sqrt{|{\bf r}_{e,v,i} -{\bf r}_{h,v',i'}|^2+d^2}},
 \label{H}
\end{eqnarray}
where terms with both $v'=v$ and $i'=i$ are excluded from  the primed sums and ${\bf r}_{e,v,i}$ ( ${\bf r}_{h,v,i}$) 
is the in-plane position of the  $i$th electron (hole) in valley $v$. Above, $m_b$ is the band effective mass of the 
carriers, which move in a medium of dielectric constant  $\epsilon$. QMC simulations of the $g_v=1$ case are  
already available both at $T=0$ \cite{depalo2002,depalo2003,maezono2013,saini2016,pablo2018} and at $T>0$ 
\cite{bonitz2012}

At given valley degeneracy, the properties of the system depend on  the interlayer distance $d$  and  the in-layer 
coupling parameter $r_s=1/\sqrt{\pi n}$, while the ratio $r_s/d$ measures the importance of the interlayer 
attraction, as compared to  the in-layer repulsion. Provided that screening is not too strong, i.e., $r_s\gtrsim1$,   a 
paired phase is expected for $r_s/d>1$ \cite{depalo2002,maezono2013}. Moreover, for $r_s/d\gg 1$ 
quadriexcitons \cite{kittel1972} should appear, instead of the  biexcitons found in the one valley case 
\cite{maezono2013}.  
In this Letter we systematically investigate the region $r_s<8$, for  systems with $N=84$ particles per layer. We 
also study some  systems with a larger number of particles, up to $N=148$,  to assess size effects, as well as 
some systems at  lower densities (larger $r_s$ values). 

We have employed variational and diffusion Monte Carlo \cite{daviddmc,cyrusdmc,rmpfoulkes} (VMC and DMC) 
as implemented in our own code.  At each  $r_s$  and $d$,  an optimal trial function $\Psi_T$ is determined by 
minimizing the  variational energy with respect to a number of optimizable  parameters \cite{linear}. We then  
compute  estimates of the properties of interest using Monte Carlo integration with $|\Psi_T|^2$ as {\em 
importance function} and, in most cases, using also the more accurate fixed-node DMC\cite{ daviddmc,cyrusdmc} 
with $\Psi_T$ as guiding function.  We have used a singlet BCS-Jastrow trial function \cite{bouchaud1988,depalo2002}
\begin{eqnarray}
\Psi_T  &=& J\prod_{\sigma}D^{\sigma,\sigma}_{e,h}, \\
D_{e,h}^{\sigma,\sigma}&=&\det[\phi({\bf r}_{e, \sigma,i}-{\bf r}_{h,  \sigma,j})],\label{BCS}
\end{eqnarray}
with $\sigma=(v,s_z)$ the valley-spin index or flavor,  and  the Jastrow factor 
\begin{equation}
J=\exp[-(1/2)\sum_{\mu,\mu'}\sideset{}{'}\sum_{i_{\mu},j_{\mu'}}
u_{\mu,\mu'}(|{\bf r}_{i_{\mu}}-{\bf r}_{j_{\mu'}}|)]
\label{Ja}
\end{equation}
embodying  two-body correlations. Above, the {\it species} index   $\mu=(t,\sigma)$ combines the particle type 
($t=e,h$) and the flavor;  moreover,  the primed sum for $\mu'=\mu$ contains only the terms with $i_{\mu}\ne 
j_{\mu}$.    The pairing orbital $\phi({\bf r})$ is chosen of a flexible form suggested earlier 
\cite{carlson2003,maezono2013},  
\begin{equation}
\phi({\bf r}) =c(r) +\sum_{l=1}^{N_k} p_{|{\bf k}_l|}\cos({\bf k}_l\cdot{\bf r}),
\label{phi}
\end{equation}
where  $c(r)$ is a spherical function of finite range $r_c \leq L/2$,  $L$ is the side of the periodic square 
simulation box,  and the sum is over closed shells of the shortest reciprocal space  wavevectors. 
The BCS part of the wavefunction $\prod_{\sigma} D_{e,h}^{\sigma,\sigma}$ is able to describe different 
homogeneous fluid phases \cite{maezono2013}. When  $ p_{|{\bf k}_l|}=0$ for all $l$  one obtains a fluid of  
excitons. On the contrary  when   $c(r)=0$, $N_k=N/4$ and $ p_{|{\bf k}_l|}\neq 0$ for all $l$ one obtains a 
plasma phase described by  plane-wave Slater determinants. In  this latter case, in fact,   
$D_{e,h}^{\sigma,\sigma}=D^{\sigma}_e 
D^{\sigma}_h$  \cite{bouchaud1988} and $D^{\sigma}_t =\det[\exp(i {\bf k}_l\cdot{\bf r}_{t, 
\sigma,i})]$. The Jastrow factor, embodying two-body terms, apart from generally improving the wavefunction, is 
crucial in making possible polyexcitonic phases\cite{pseudo}.

The function   $c(r)$ and  all pseudopotentials  $u_{\mu,\mu'}( r)$ in the Jastrow factor 
are expanded on a flexible basis of locally piecewise-quintic Hermite interpolants \cite{natoli95}, which among 
other things easily accommodates constraints at the end points. For each function its radial range and the 
expansion coefficients provide the variational parameters.  For the pairing orbital $\phi(r)$ such a set of variational 
parameter is augmented   by the plane wave coefficients $ p_{|{\bf k}_l|}$.  Depending on the chosen numbers of 
knots in  the radial mesh and plane wave coefficients, the overall number of  variational parameters used in the 
simulations describes below is typically between about 50 and 60. \cite{sup19}

\begin{figure}[]
  \includegraphics[width=0.50\textwidth, trim = 20 15 0 0]{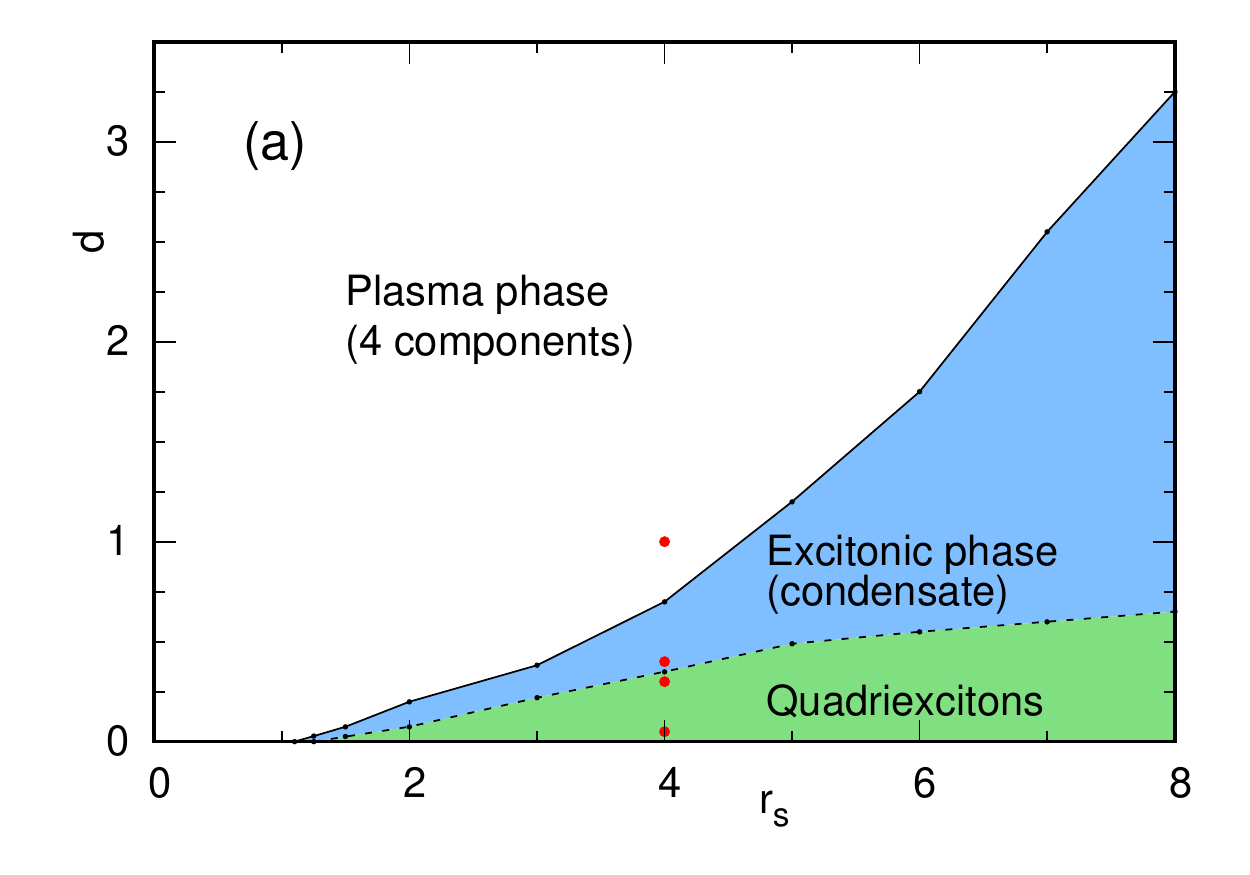}
  \par
  \vspace{1mm}
  \includegraphics[width=0.5\textwidth, trim = 20 15 0 0]{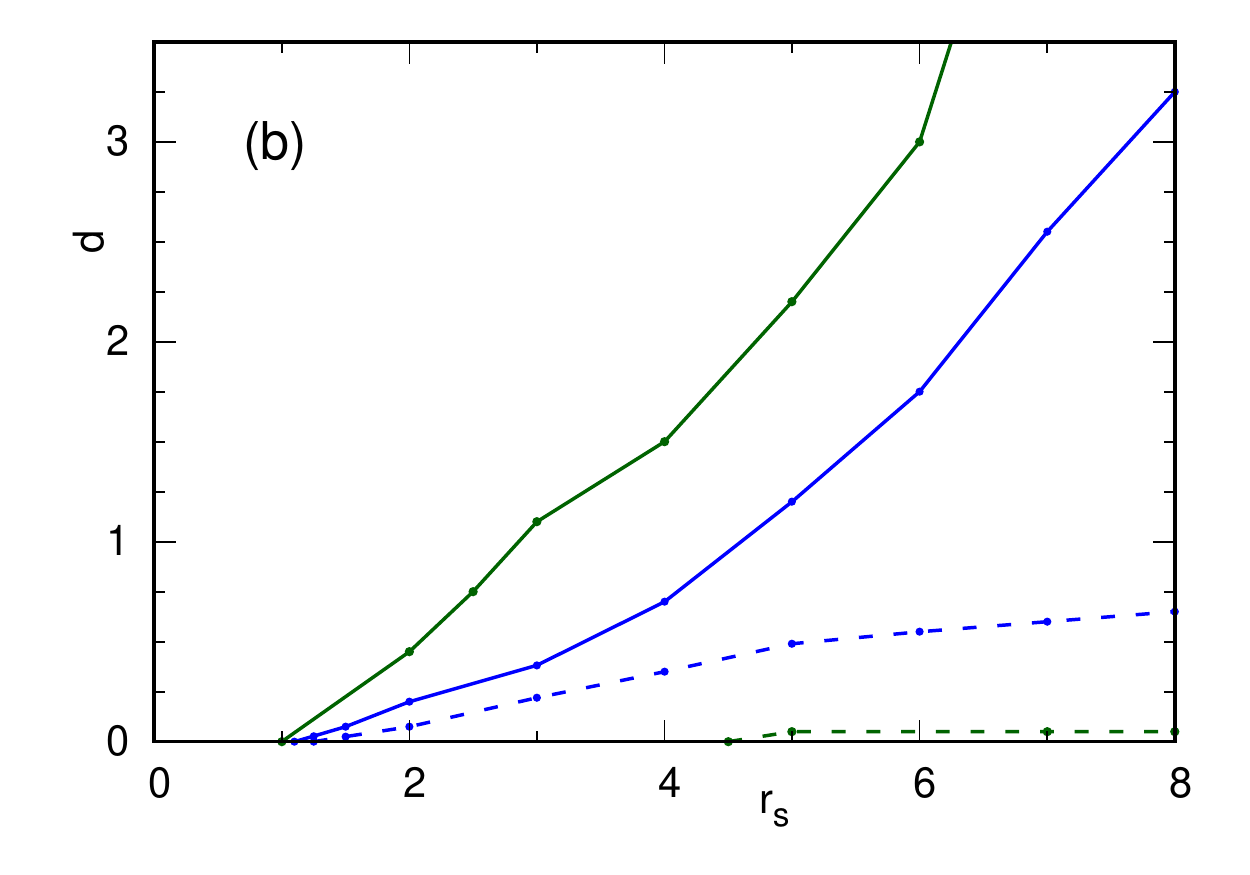}
\caption{(a) Phase diagram of an electron-hole bilayer with twofold  valley degeneracy for small to moderate values 
of the intralayer coupling $r_s$ and interlayer distance $d$. The red dots indicate states that are studied in Figs. 
\ref{plateau} and \ref{geh}. The regions of stability of the competing (fluid) phases are displayed in different 
colors.  (b) Comparison of the phase diagrams of electron-hole bilayers  with one and two valleys. The region of 
stability of the excitonic  condensate  in the one-valley system (area between the green lines  \cite{maezono2013,phdia1v} )  
is substantially reduced in going to the two-valley system (area between the blue lines). } 
  \label{phase-diagram}
\end{figure}

\begin{figure}[]
 \includegraphics[width=0.5\textwidth, trim = 20 5 0 0]{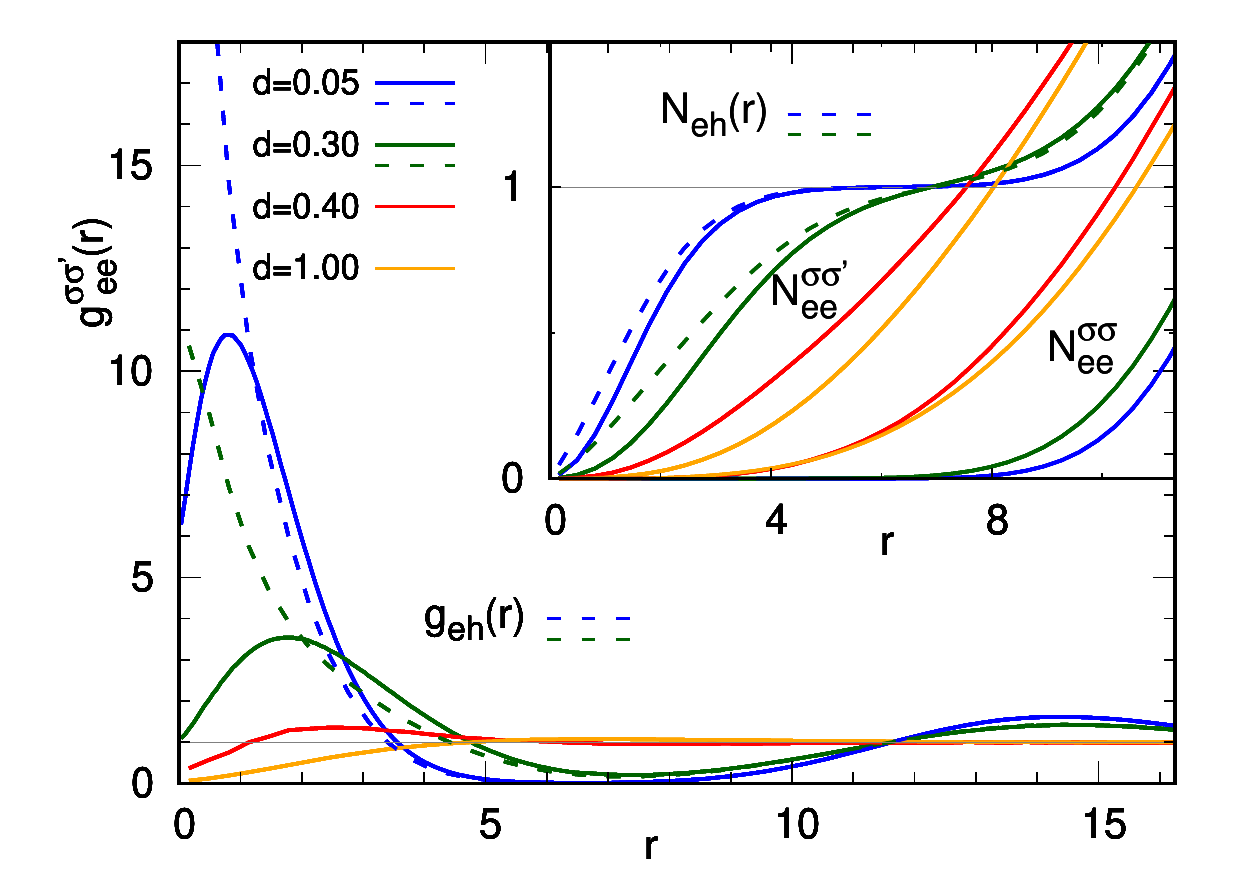}
\caption{PCFs (full lines) between electrons (holes) with different flavors ($\sigma'\ne \sigma$) in a symmetric 
electron-hole bilayer with twofold valley degeneracy, at $r_s=4$;  interlayer distances $d$ correspond to various 
phases (see Fig \ref{phase-diagram}).  The inset reports  the RCNs 
  defined in  Eq. \ref{coordination-diff}. Electron-hole  PCFs and RCNs are also shown for the two smaller 
distances (dashed lines), where they turn out to be independent of flavors, which have therefore been dropped. }
  \label{plateau}
\end{figure}

The main outcomes of our DMC simulations are summarized in  the phase diagram displayed in 
Fig.\@\ref{phase-diagram}(a).  For small values of the intralayer coupling $r_s\lesssim1$ (large density),  a 
four-component plasma phase is found  stable at all distances evidently due to strong screening. As one turns to  
larger values of $r_s$ ($r_s\gtrsim 1$)  the interlayer attraction becomes more effective and stabilizes a novel 
quadriexcitonic phase at  smaller  distances and an excitonic phase with condensate at intermediate distances. 
As illustrated in  Fig.\@\ref{phase-diagram}(b), with respect to the findings of  DMC simulations for a symmetric  
electron-hole bilayer without valley degeneracy 
\cite{depalo2002,maezono2013} we obtain a substantial shrinking of the region of stability of the excitonic phase, 
especially at  small $r_s$. 
This may partly explain the difficulties encountered in finding experimental evidence of excitonic condensation in 
coupled electron-hole graphene bilayers \cite{deanGBprl,tutucGBprl,tutucprl2018} 
and suggests  avoiding high density in the search of the condensate, as has been noted before \cite{peralietal13}.  
In the density range investigated here, 
we find no evidence of   biexcitons\cite{maezono2013}, which appear to be substituted by the much more stable 
quadriexcitons \cite{threexcitons}. Below we characterize  the various phases by analyzing  features of the extrapolated 
estimates of pair correlation functions  and  condensate fraction \cite{sup19}.

In Fig. \ref{plateau} we report the  pair correlation function (PCF) \cite{sup19}  between electrons with different 
flavor $g_{ee}^{\sigma\sigma'}(r)$, $\sigma'\ne \sigma$,  at $r_s=4$. By symmetry, 
$g_{ee}^{\sigma\sigma'}(r)=g_{hh}^{\sigma\sigma'}(r)$. 
In the plasma phase ($d=1$) the PCF is  structureless for $r \gtrsim 5 $, with a 
modest correlation hole for  $r\lesssim 5$. In the excitonic phase ($d=0.4$)  the PCF remains structureless at 
larger distances, though showing a modest  increase  at smaller distances. As one crosses into the 
quadriexcitonic phase ($d=0.3$) a large peak (higher than 3 ) appears at  small distances, followed by a deep  
wide minimum, inducing  pronounced oscillations at large distances. This behavior,    is further enhanced at 
$d=0.05$ with a peak as large as 11 and a minimum of 0.  So, what is causing an "effective attraction" between 
unlike flavor electrons (holes), that overcomes direct Coulomb repulsion and produces such a short-range 
ordering? We note that this phenomenon is accompanied by the fact that  the electron-hole PCF 
$g_{eh}^{\sigma\sigma'}(r)$,  which is strongly peaked on the electron, become independent on the flavor 
and  display the same deep wide  minimum found in $g_{ee}^{\sigma\sigma'}(r)$, $\sigma'\ne \sigma$. A clue to 
what is going on is provided by the inspection of the distance-dependent pile-up of  particles of the species 
$(t',\sigma')$ around a particle of the species $(t,\sigma)$, i.e., the  running coordination number (RCN)  
\begin{eqnarray}
\label{coordination-diff}
N_{ tt'}^{\sigma\sigma'}(r)&=&2\pi n_{t',\sigma'}\int_0^r\, ds\, s\, g_{tt'}^{\sigma\sigma'}(s)
\label{coordination-eq},
\end{eqnarray}
$n_{t,\sigma}$ being a species areal density.

In the inset of Fig. \ref{plateau} we display $N_{ee}^{\sigma\sigma'}(r)$ and $N_{eh}^{\sigma\sigma'}(r)$, 
respectively counting the average number of electrons (holes) with  flavor  $\sigma'$ around an electron with 
flavor $\sigma$.  Let us consider the case $d=0.05$ first.  It is evident that electrons with the same flavor  as the 
one at the origin  are completely expelled from a very large region, $N_{ee}^{\sigma\sigma}(r) \simeq 0, \, 
r\lesssim 8$. They make space for a dynamic compound of radius about 5, comprising 4 electrons (the one at 
the origin and 3 with $\sigma'\ne \sigma$)  and 4 holes of the four flavors. This is what we call quadriexciton.  At 
$d=0.05$ the quadriexciton appears to be well defined,  being neatly separated by its nearest neighbors, 
whereas at $d=0.30$ neighboring quadriexcitons  touch and at $d=0.40$ they are melted into excitons.

\begin{figure}[]
\includegraphics[width=0.5\textwidth, trim = 20 5 0 0]{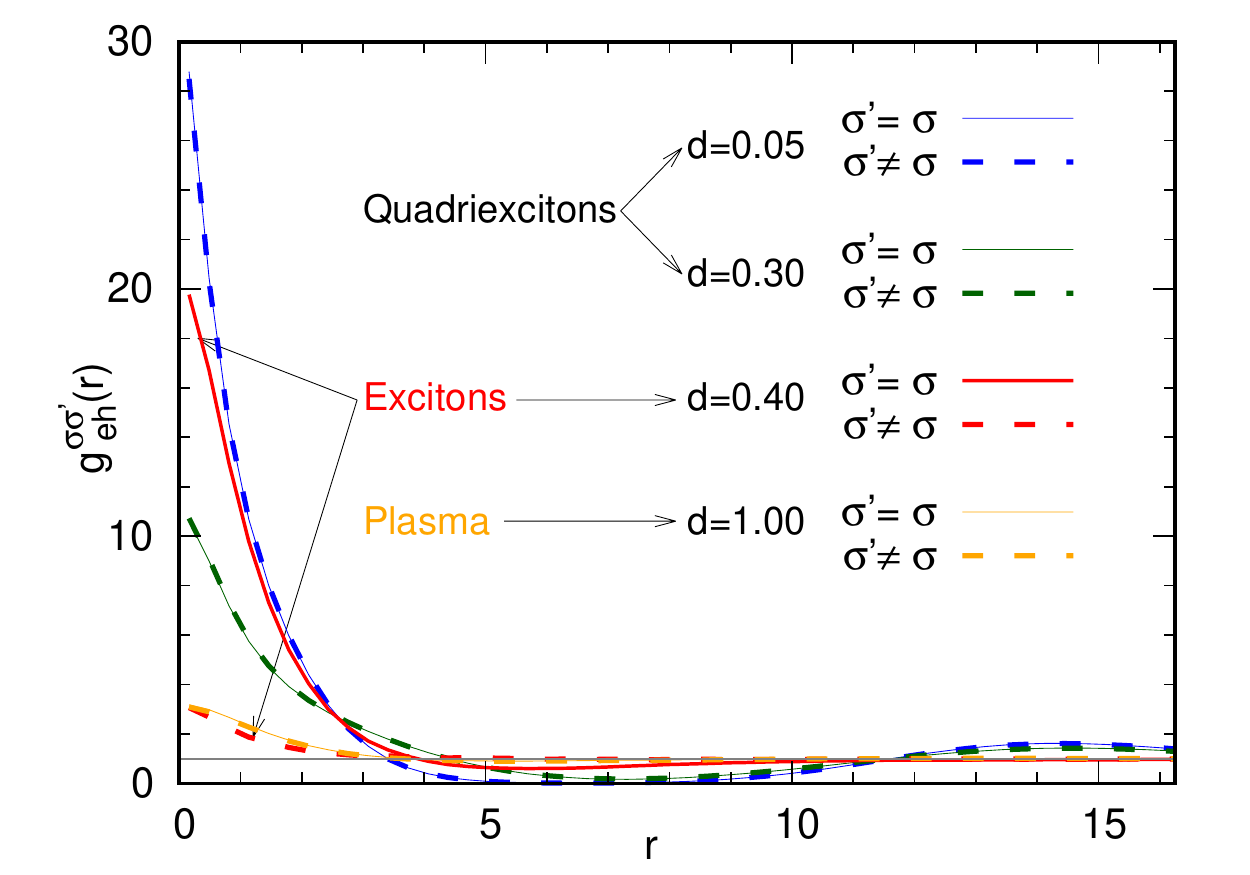}
\caption{Electron-hole PCFs in a symmetric electron-hole bilayer with twofold valley degeneracy, at $r_s=4$ and 
interlayer distances $d$ corresponding to various phases (see Fig \ref{phase-diagram}). The PCFs of 
quadriexcitons and plasma are independent  of flavors $\sigma,\sigma'$.}
  \label{geh}
\end{figure}

The comparison in Fig. \ref{geh} of the electron-hole PCFs of the various phases at $r_s=4 $   reveals   in the 
excitonic phase a large  difference between   paired ($\sigma'=\sigma$)  and unpaired ($\sigma'\ne \sigma$) 
$g_{eh}^{\sigma\sigma'}(r)$, the height at the origin differing by  as much  as  a factor 6. On the contrary, 
$g_{eh}^{\sigma\sigma'}(r)$ is flavor independent  both in the plasma phase, a consequence of the symmetry 
of the wavefunction, and in the quadriexcitonic phase,  seemingly   as  an effect  of  the interplay    of 
pseudopotentials and  paring orbital in the energy minimization. Such an interplay  often results in  a very 
repulsive pseudopotential  $u_{\mu,\mu} (r)$ between electrons (holes) with the same flavor, as found also in 
the biexcitonic phase   of the one-valley system \cite{maezono2013}. We  remark that the qualitative features of 
PCFs and RCNs of the various phases illustrated above for $r_s=4$  are common to the whole density range  
displayed in  Fig.\@\ref{phase-diagram}.    

\begin{figure}[]
 \includegraphics[width=.5\textwidth,trim = 20 5 0 0]{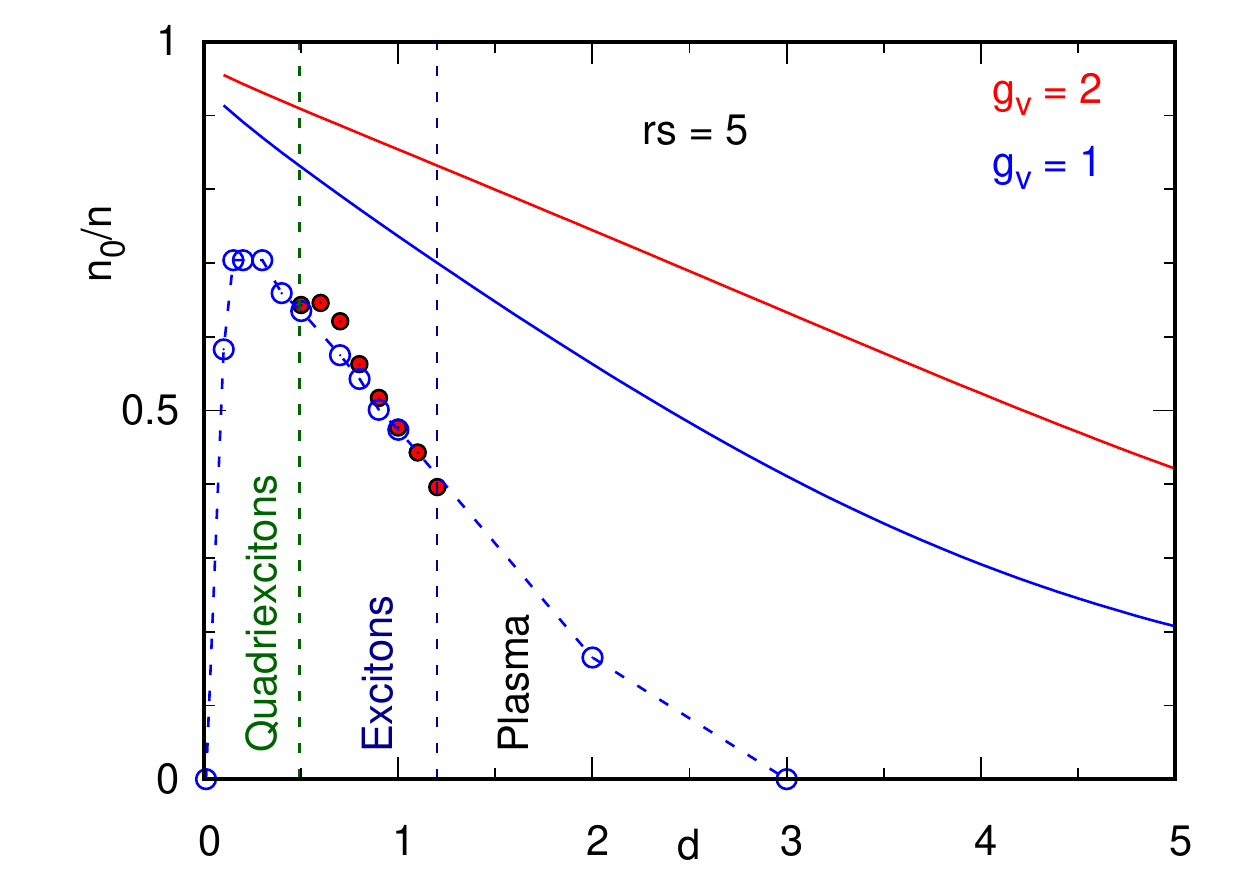}
\caption{ Condensate fraction versus  interlayer distance $d$ in a symmetric electron-hole bilayer, at $r_s=5$: (a) present results
 for $g_v=2$ (red dots), with  vertical dashed lines  providing boundaries between phases;  (b) $g_v=1$ \cite{maezono2013}  
 (blue circles ), with the dashed blue line a guide to the eyes; (c)  BCS mean field \cite{prllittlewood1995} predictions (full curves).}
 \label{cond-frac}
\end{figure}

The condensate fraction $n_0/n$ in the excitonic phase is shown in Fig. \ref{cond-frac} for $r_s=5$, as function of 
the interlayer distance $d$ \cite{sup19}. Comparison of the present results with those for the one-valley system  
\cite{maezono2013} reveals that valley degeneracy, while substantially reducing the region of stability of the 
excitonic phase, leaves essentially unchanged the value of the condensate fraction. On the other hand,  
comparison with the predictions of a BCS treatment  \cite{prllittlewood1995}, uncovers the limitations of such a  
mean-field approach, which predicts for the condensate a substantial increase  with valley degeneracy and a 
much slower decay  with the distance $d$ . The results at $r_s=5$ are  representative of our findings in the full 
range of density investigated in the present work, i.e.,  $ r_s\leq 8$, when one keeps in mind that (i)  the  stability 
window  (in distance) of the excitonic phase  grows with $r_s$, as it is clear from Fig. \ref{phase-diagram},  and 
(ii) the condensate fraction  becomes somewhat smaller than in the one-valley system at large densities 
($r_s<3$).

The  excitonic condensate  at $T=0$ may be found in three regimes \cite{leggett1980}:   BCS   at high density, BEC  
at low density and BCS-BEC crossover  in the middle.  In order to characterize  the regime for given $d$ and $r_s$  
one may use  the  values of  the condensate fraction $n_0/n$  \cite{guidini2014,pablo2018} and $k_Fr_{ex}$ \cite{pistolesi1993}, 
the ratio     of the exciton  radius $r_{ex}$  and the interexciton distance $\sim 1/k_F$, $k_F$ being  the Fermi 
wavevector of electrons. The exciton radius is  evaluated as 
\begin{equation}
 r_{ex}^2=2\pi n_{h,\sigma}\int_0^{r_1}\, ds\, s\, g_{eh}^{\sigma\sigma}(s)s^2,
\end{equation} 
with $r_1$ the radius of the circle centred on an electron containing on average one hole.
In the density range studied here we find \cite{sup19}   $  0.21 \le  n_0/n \le  0.75$ and $0.31\le k_Fr_{ex}\le 0.96$ which,  
using the criteria  of Refs.   \onlinecite{guidini2014,pablo2018}, places the condensate always in the   BEC-BCS crossover regime.

We now turn  to a peculiar aspect of the determination of the phase boundaries in  Fig.\@\ref{phase-diagram}. 
The  wavefunctions obtained through energy minimization display an hysteresis phenomenon when crossing a 
boundary. For the sake of clarity consider the quadriexciton to exciton transition at fixed $r_s$, which takes places 
at $d_{qe}(r_s)$ and choose $r_s=4$, where  $d_{qe}(4) =0.35$.  If at $d\gtrsim 0.35$ we start the energy 
minimization with a converged quadriexcitonic wavefunction taken from $d<0.35$ the wavefunction remains 
quadriexcitonic in character, with zero condensate. However, by increasing $d$ further  the wavefunction 
eventually becomes excitonic, with a finite condensate. Similarly, if at $d\lesssim 0.35$ we start with a converged  
excitonic wavefunction taken from $d>0.35$  we find that the wavefunction remains excitonic; however if we 
decrease $d$ further the wavefunction eventually becomes quadriexcitonic, with zero condensate. This implies 
that in the vicinity of  $d=0.35$  we have two solutions with different character and different VMC energies as well 
as  fixed-node DMC energies. The latter more accurate energies  are thus used to determine the stable phase 
near the boundary\cite{sup19}. 

To conclude, we comment on the correspondence of the calculated phase diagram  (Figs. 1) with interlayer conductance 
measurements in double-bilayer graphene-WSe$_2$ heterostructures\cite{tutucprl2018}. In the experiment, a nominally 
divergent differential conductance between the  two graphene bilayers, observed in a density interval $(n_{\rm min},n_{\rm max})$, 
is attributed to condensation of electron-hole pairs\cite{west2000,tutucprl2018,faimak2019}. Suppression of condensation 
is ascribed to in-plane  screening for $n>n_{\rm max}$ and to disorder or competing phases for $n<n_{\rm min}$, as generally expected 
for indirect excitons.\cite{comte1982,littlewood1996}.We can be more specific and relate the onset of pair 
condensation to phase transitions from plasma to excitons on the high density side, which is common to 
one-valley electron-hole devices\cite{maezono2013,faimak2019}, and from quadriexcitons to excitons on the low 
density side, which represents a mechanism peculiar to two-valley devices even in the absence of disorder\cite{nota}. 
We note that the large extent of the quadriexcitonic phase, compared to the biexcitonic phase  in  the one-valley 
system\cite{maezono2013}, is instrumental  in having suppression of pair condensation at low  density for interlayer  spacings 
accessible experimentally.

The experimental observation of condensation is strongly peaked around conditions
of balanced electron and hole densities\cite{tutucprl2018}, matching the symmetric two-valley
system considered here. Other aspects of our model, such as the assumption of parabolic bands\cite{conti2019}
and isotropic dielectric constant, are less faithfully representative of the actual 
heterostructure. Nevertheless, using the largest and the smallest dielectric constants\cite{tutucprl2018}
of the constituent materials to translate device parameters to our units,
the experimental spacing between graphene bilayers varies from 0.1 to 0.5. This is well
within the region where we find a non-zero condensate bracketed by the plasma and
the quadriexciton phases. 

Our results support and complement measurements of interlayer tunneling  
conductance\cite{west2000,tutucprl2018,faimak2019} as a probe for indirect exciton condensation.


\begin{thebibliography}{57}
\bibitem [{lin()}]{linear}%
  \BibitemOpen
  \href@noop {} {}\bibinfo {note} {We have used the linear method
  \cite{linearmethod,linearmethod1} and checked, in selected cases, that the
  obtained minimum agrees with the one found by the improved stochastic
  reconfiguration method \cite{sorella-sr}}\BibitemShut {NoStop}%
\bibitem [{pse()}]{pseudo}%
  \BibitemOpen
  \href@noop {} {}\bibinfo {note} {We have used 4 independent pseudopotentials
  $u_{\mu,\mu'}$, with appropriate cusp conditions \cite{kato}: same-flavor
  same-particle pairs, different-flavor same-particle pairs, same-flavor
  electron-hole pairs, and different-flavor electron-hole pairs. $\phi(r)$ is
  taken cuspless.}\BibitemShut {Stop}%
\bibitem [{not()}]{nota}%
  \BibitemOpen
  \href@noop {} {}\bibinfo {note} {Quadriexciton condensation is also possible
  in principle, but we expect that enhanced mass, reduced density and
  interaction-induced depletion will strongly suppress it relative to pair
  condensation.}\BibitemShut {Stop}%
\end{thebibliography}
%
\end{document}


\title{Quadriexcitons and excitonic condensate in a symmetric electron-hole bilayer with valley degeneracy: Supplemental Material }

\author{Stefania De Palo}
\altaffiliation{depalo@iom.cnr.it}
\affiliation{CNR-IOM-DEMOCRITOS, Trieste, Italy}
\affiliation{Dipartimento di Fisica,
Universit\`a di Trieste, strada Costiera 11, 34151
Trieste, Italy}

\author{F. Tramonto}    
\altaffiliation[Present address: ]{Kyndryl Italia Innovation Services S.r.l., Via Circonvallazione Idroscalo snc, 20090 Segrate (MI),  Italy}\affiliation{Dipartimento di Fisica,
Universit\`a di Trieste, strada Costiera 11, 34151
Trieste, Italy}

\author{Saverio Moroni}
\altaffiliation{moroni@iom.cnr.it}
\affiliation{CNR-IOM-DEMOCRITOS, Trieste, Italy}
\affiliation{SISSA (International School for Advanced 
Studies), via Bonomea 265, 34136 Trieste, Italy}

\author{Gaetano Senatore}
\altaffiliation{senatore@units.it}
\affiliation{Dipartimento di Fisica,
Universit\`a di Trieste, strada Costiera 11, 34151
Trieste, Italy}

\maketitle

\renewcommand{\thetable}{S\arabic{table}}
\setcounter{figure}{0}
\renewcommand{\thefigure}{S\arabic{figure}}
\setcounter{section}{0}
\renewcommand{\thesection}{S\arabic{section}}
\setcounter{equation}{0}
\renewcommand{\theequation}{S\arabic{equation}}
\renewcommand{\bibnumfmt}[1]{[S#1]}
\renewcommand{\citenumfont}[1]{S#1}

This Supplemental Material contains information on (i) details of the
simulations, (ii) optimization of the wave function across a phase boundary,
(iii) estimate of the condensate fraction, and (iv) structure of the
quadriexcitons. Unless otherwise specified, energies are in Ry$^*$
and distances in $a_B^*$.

\section{Details of the simulations}
The calculations presented in the main text have been done with $N=168$
particles  \cite{remark} in periodic boundary conditions. The DMC energies, used to map
the phase diagram, are calculated with a number of walkers 
$N_W=440$ and a time step $\tau$ tuned to achieve about 90\% acceptance rate 
for global moves. The biases due to finite $N$, $N_W$ and $\tau$,
estimated in a few representative cases, are all negligible for our
purposes, as we show below.
\begin{figure}[b]
 \includegraphics[width=0.50\textwidth, trim=0 0 0 0]{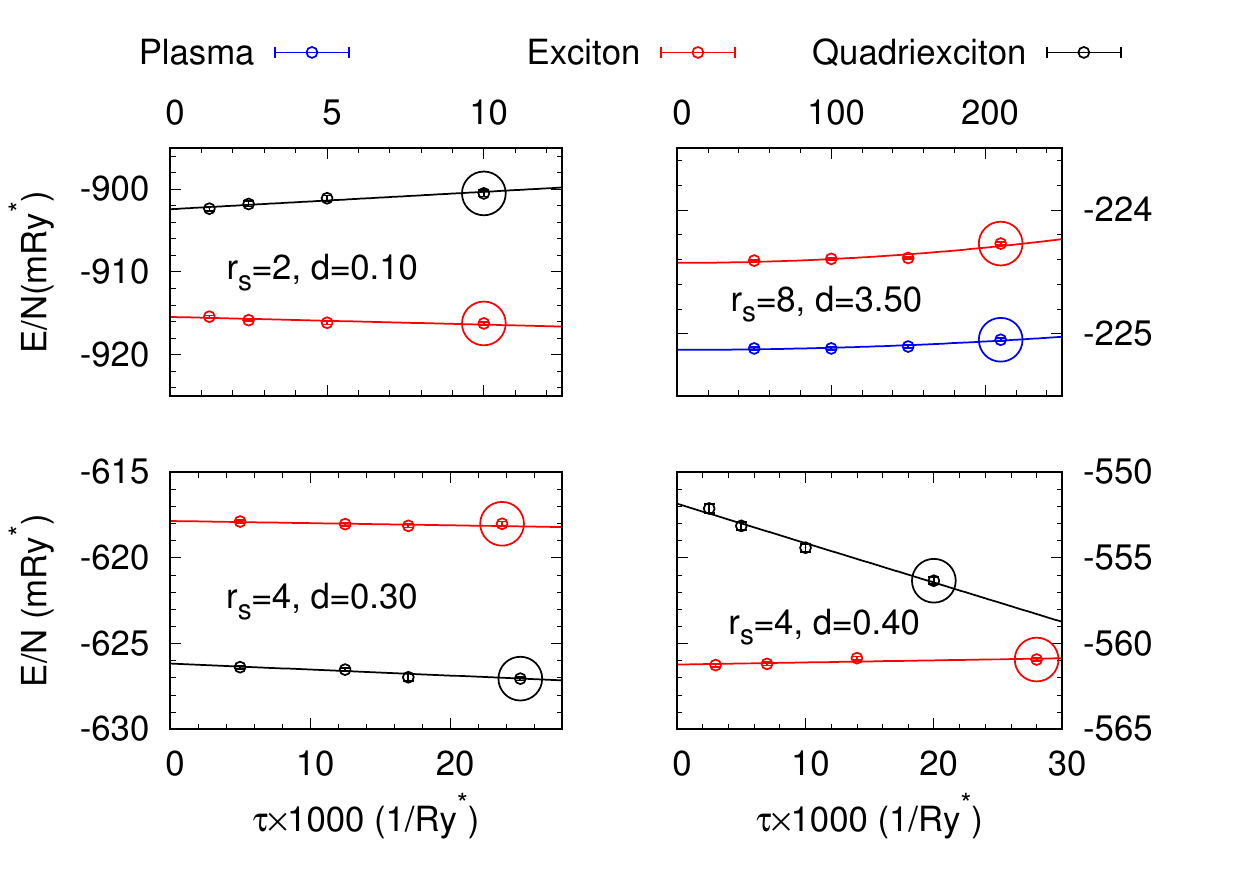}
\caption{Dependence of the DMC energy on the time step $\tau$ for
various phases at selected points of the phase diagram, as specified 
in the body of each panel.
}
  \label{tstep}
\end{figure}

\subsection{Time step error}
The criterion of 90\% acceptance rate for the choice of $\tau$ affords
a rather uniform accuracy across various densities, interlayer distances,
and phases, without the computational burden of explicit extrapolations
to zero time step. We have verified in selected cases, some of
which are shown in Fig.~\ref{tstep}, that the time step bias incurred 
by this choice does not affect our phase diagram in a significant way.

We address in detail the more problematic case we encountered,
the exciton-quadriexciton transition at $r_s=4$ between $d=0.3$ and $d=0.4$,
inferred from the data shown in the lower panels of Fig.~\ref{tstep}. 
In particular, the energy of the quadriexciton phase at $d=0.4$ (black
symbols in the bottom right panel) has an unusually large time step error
of nearly 5mRy$^*$.
The phase boundary is defined as the zero of the linear interpolation
between $d=0.3$ and $d=0.4$ of the energy difference between the two 
phases, $\Delta e(d)\equiv (E_{\rm exc}(d)-E_{\rm qua}(d))/N$. 
The data taken at 90\% acceptance rate, marked by
circles, yield an energy difference (in mRy$^*$) $\Delta e(0.3)=9.0$
and $\Delta e(0.4)=-5.1$, while the extrapolations
at $\tau=0$ yield $\Delta e(0.3)=8.3$ and 
$\Delta e(0.4)=-9.4$, all with statistical uncertainties of 0.3 or less.
The effect of the time step extrapolation is a downward shift of the
boundary by no more than 0.02, which would be barely visible on the 
phase diagram shown in the main text.

\subsection{Finite size error}
\begin{figure}[]
 \includegraphics[width=0.5\textwidth, trim=0 0 0 0]{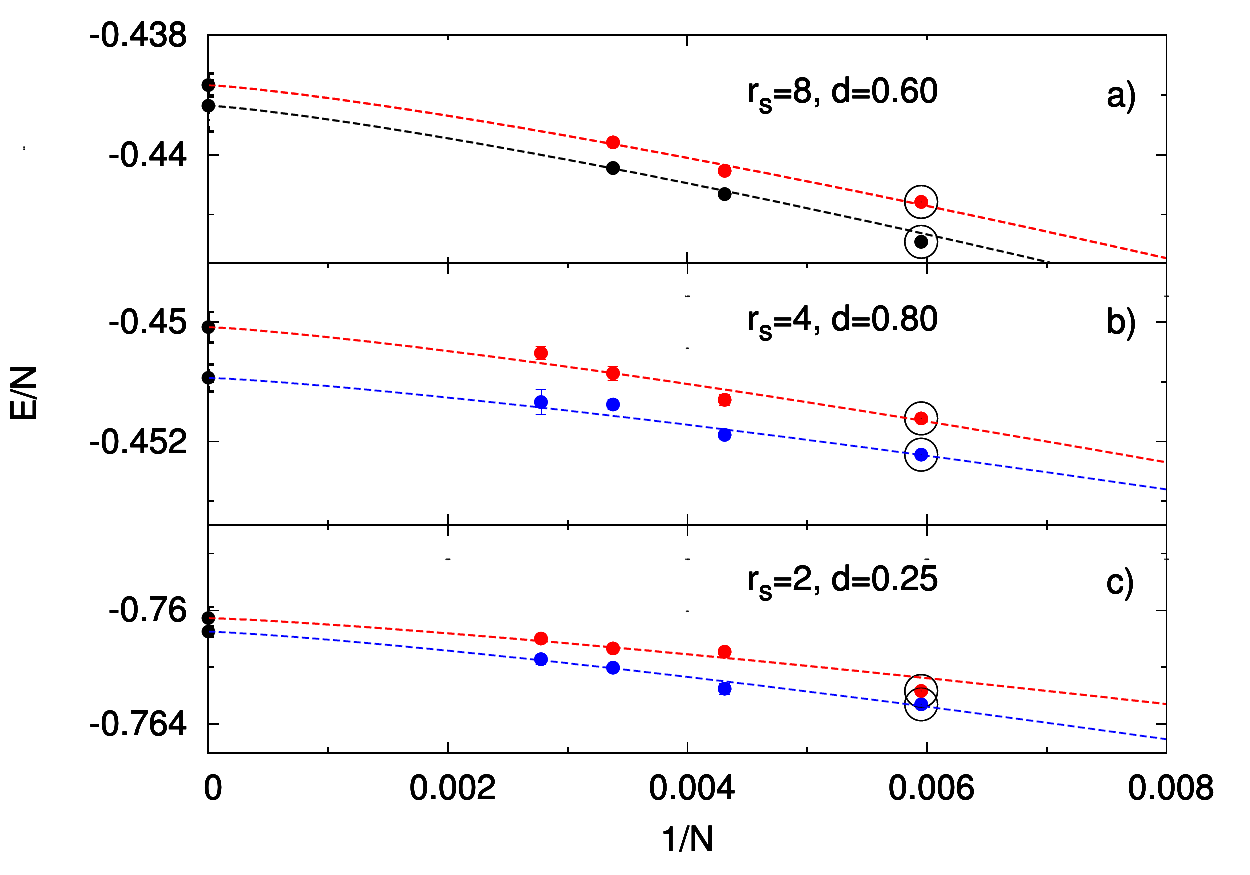}
\caption{Dependence of the DMC energy of various phases on the number 
of particles $N$, at selected points of the phase diagram specified 
in the body of each panel.
Blue, red and black symbols refer to the plasma, exciton and quadriexciton
phase, respectively (the color code does not apply to the extrapolated
values, shown in black for all phases).
For the plasma phase, the ideal gas finite-size correction has been subtracted.
The data used in the main text, corresponding to $N=168$, are circled.
The other system size considered are $N$=232 and 296. In panels b) and
c) we add data for $N=320$ because they involve the plasma phase
whose size extrapolation may be more problematic due to residual
shell effects. All energies are extrapolated to zero time step.
}
  \label{size}
\end{figure}

Finite-size shell effects in the plasma phase are reduced
by subtracting the ideal gas finite-size correction. Smooth finite size 
effects, with a leading correction\cite{foulkes} proportional to $N^{-5/4}$,
are expected to be similar for the various phases,
and to cancel out to a large extent in the energy differences used to
map the phase diagram. This is verified in Fig.~\ref{size}, where we 
show explicit size extrapolations to the thermodynamic limit of DMC 
energies in selected cases near phase boundaries. 
The finite size error at $N=168$ is few mRy$^*$ for the energy of any 
given phase, and a fraction of mRy$^*$ for the energy difference between
phases. Its effect on the phase diagram is completely negligible.

\subsection{Population control bias}
\begin{figure}[]
 \includegraphics[width=0.50\textwidth, trim=0 0 0 0]{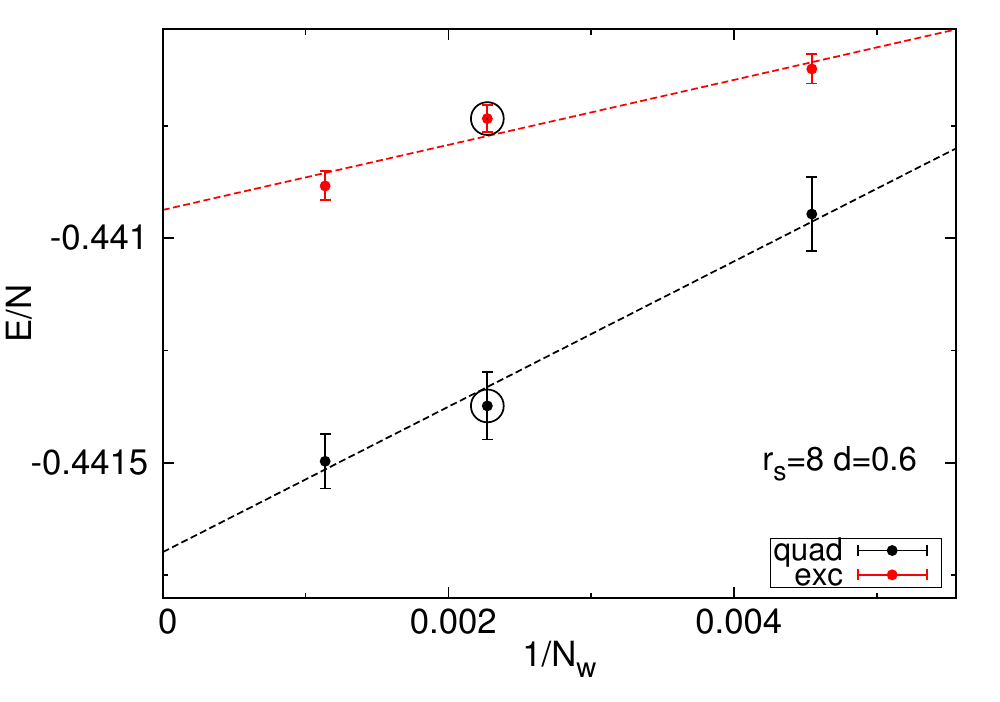}
\caption{Extrapolation of the DMC energy to infinite number of walkers for the
exciton and quadriexciton phases at $r_s=8, d=0.6$. 
The data used in the main text are circled.
}
  \label{fig_nw}
\end{figure}

The number of walkers in our DMC simulations is $N_W=440$. The population 
control bias a fraction of mRy$^*$ for the energy of a given phase,
and even smaller for the energy difference between phases, as shown
in the extrapolation to $1/N_W\to\infty$ illustrated in Fig.~\ref{fig_nw}. 
Again, its effect on the phase diagram is completely negligible.

\section{Optimizing across phase boundaries, and mapping the phase diagram}

The functional form of our wave function has the flexibility to
represent the quadriexciton, the exciton, and in principle the 
plasma phases (although for the latter we use in practice a Slater 
determinant, to which the s-wave pairing function reduces as the 
Fourier transform of the pair orbital approaches the Heaviside function 
$H(k_F-|k|)$).

\begin{figure}[]
 \includegraphics[width=0.5\textwidth, trim=0 0 0 0]{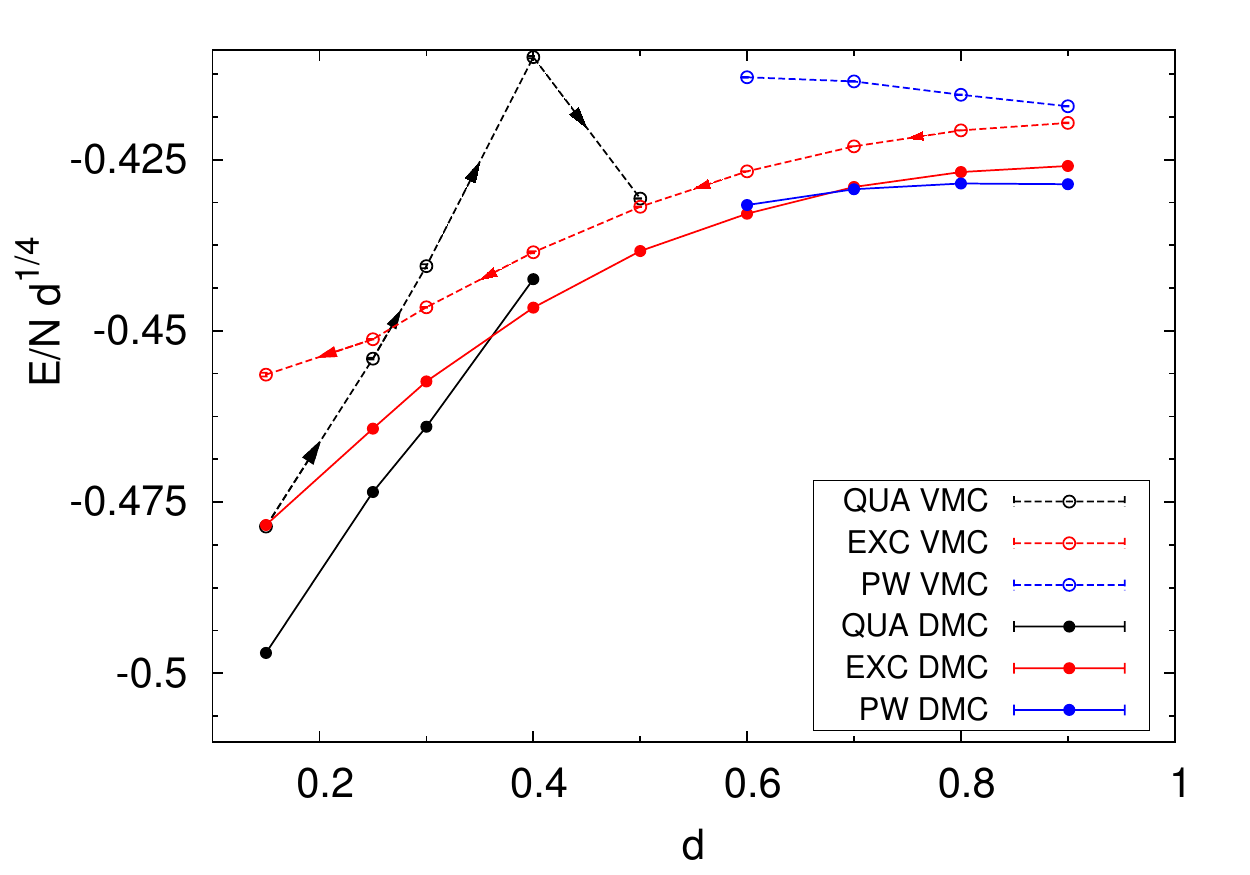}
\caption{Energy as a function of the interlayer distance $d$ for
the plasma (blue), exciton (red) and quadriexciton (black) phases 
at $r_s=4$, calculated with VMC (open circles, dashed lines) and DMC 
(filled circles, solid lines). The arrows indicate the direction of the
optimization series (see text).
}  \label{hysteresis}
\end{figure}

An example is shown in Fig.~\ref{hysteresis}, where the empty circles
joined by dashed lines show the VMC energies at $r_s=4$ as a function 
of $d$. The black symbols represent a series of optimizations with
increasing values of $d$, as indicated by the arrows, starting in the 
quadriexcitonic phase at $d=0.15$. The phase, inferred from structural 
properties as detailed in the main text and in the last section of this Supplemental Material, 
remains quadriexcitonic even for $d=0.3$ and $d=0.4$, where the VMC energy of the exciton
is lower. Eventually, for $d=0.5$, the phase collapses into excitons.
The reverse optimization series, starting in the excitonic phase at $d=0.9$
and shown by the red symbols, features a similar resilience against
phase change around $d\lesssim 0.2$. The optimization series for the plasma
phase (blue symbols) has no arrows because, as mentioned, it is done 
with a different functional form. In this case, the persistence of 
a metastable phase is trivial.

For reference, the energies displayed in Fig.~\ref{hysteresis} are listed
in Table~\ref{table_hyst}.

\begin{table*}[]
\caption{VMC and DMC energy per particle $E/N(d)$
at $r_s=4$ for the plasma phase (PW),
the excitonic phase (EXC) and the quadriexcitonic phase (QUA) 
for various distances.
DMC simulations are for $N=168$ total particles, $N_w=440$ walkers and the time steps
are to give an acceptance rate of approximately 90\%.}
\label{table_hyst}
\begin{center}
\begin{tabular}{|ccc|ccc|ccc|ccc|ccc|ccc|ccc||}
\hline
&        && &      QUA            & &&      EXC    & &&  PW        & \\
\hline
& $d$   && & VMC        & DMC     & &  VMC        &  DMC        & & VMC   &  DMC  &\\
 \hline
&0.15 && & -0.7690(2) & -0.7987(2) & & -0.7333(4)  & -0.7687(3)  & &   --- &--- & \\
&0.25 && & -0.6421(2) & -0.6270(2) & & -0.63812(6) & -0.6566(2)  & &   --- &--- & \\
&0.3  && & -0.5953(4) & -0.6270(2) & & -0.60340(6) & -0.6180(1)  & &   --- &--- & \\
&0.4  && & -0.5156(2) & -0.5563(2) & & -0.5514(2)  & -0.56156(4) & &   --- &--- &\\
&0.5  && &   ---      &   ---      & & -0.51358(4) & -0.52127(3) & &   --- &--- &\\
&0.6  && &   ---      &   ---      & & -0.48481(3) & -0.49185(2) & & -0.46922(7) & -0.49041(7) &\\
&0.7  && &   ---      &   ---      & & -0.46251(1) & -0.46898(4) & & -0.45212(7) & -0.46935(7) &\\
&0.8  && &   ---      &   ---      & & -0.44485(3) & -0.45115(4) & & -0.43935(4) & -0.45306(5) &\\
&0.9  && &   ---      &   ---      & &             &             & & -0.42832(3) & -0.44002(4) & \\
\hline
\end{tabular}
\end{center}
\end{table*}
These results expose a potentially severe ambiguity in mapping the 
phase diagram on the basis of structural properties along a directional 
optimization series. In the example of Fig.~\ref{hysteresis}, the
quadriexciton-exciton boundary would be estimated at $d>0.4$ starting
from the quadriexciton and $d<0.15$ starting from the exciton.
A better criterion is to pick the phase with lower variational energy.
For the phase diagram presented in the main text we used our best
variational upper bounds to the exact ground state energies, given
by the DMC results. In Fig.~\ref{hysteresis}, the range of 
stability of the excitonic phase estimated from VMC energies, 
$d$ from 0.27 to over 0.9, is about twice as large than our best 
result, based on DMC energies.
\begin{figure}[b]
 \includegraphics[width=0.50\textwidth, trim=0 0 0 0]{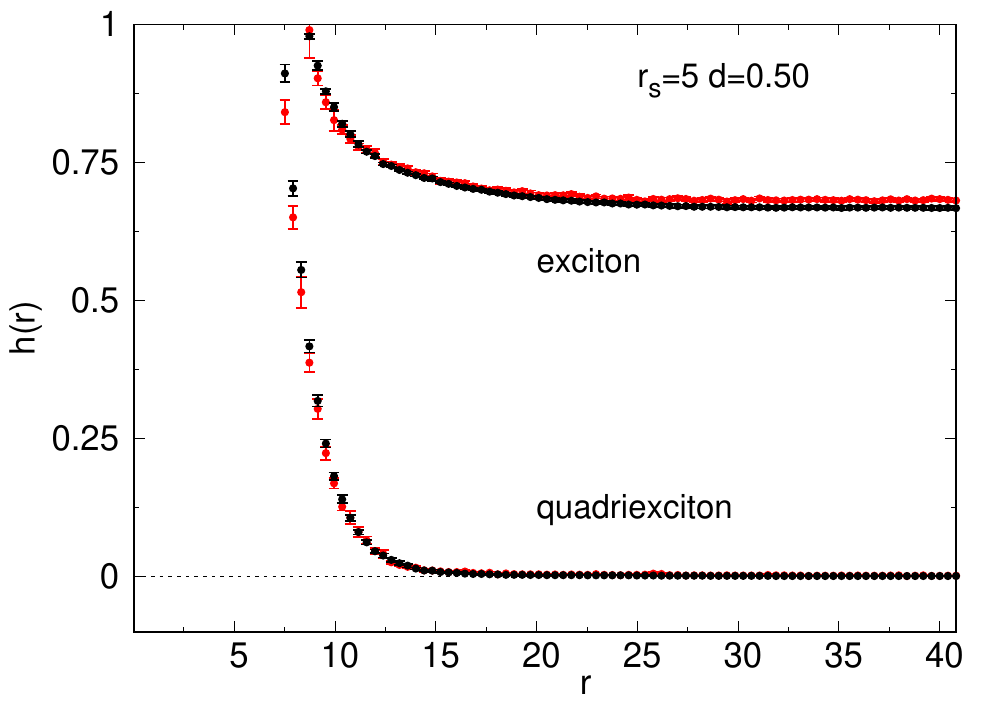}
\caption{Mixed (black) and variational (red) estimates of the reduced 
pair density matrix $h(r)$ for $r_s=5$ and $d=0.5$ for
the exciton and quadriexciton phases as indicated in the body of the figure.}
  \label{fig_hofr}
\end{figure}

\section{Condensate fraction}
The pair condensate in the exciton and quadriexciton phases is defined,
following Ref.~\onlinecite{depalo2002}, as the large-$r$ limit of the
projected two-body density matrix
\begin{equation}
h_\sigma(r)=\frac{1}{N}\int d{\bf r}_{e,\sigma}d{\bf r}_{h,\sigma}
\rho_2({\bf r}_{e,\sigma}+{\bf r},{\bf r}_{h,\sigma}+{\bf r};{\bf r}_{e,\sigma},{\bf r}_{h,\sigma})
\label{rho2}
\end{equation}
averaged over the valley-spin index $\sigma$, where $\rho_2$ is the
full two-body density matrix. 
\begin{table}[]
\caption{Extrapolated estimates \cite{tailfit} of the condensate fraction  $n_0/n$ for selected densities 
and distances in the region of stability of the excitonic phase. The uncertainty on the estimates is about  $3\%$.}
\label{table_conde}
\begin{center}
\begin{tabular}{|c||c||c||c||c||c||c|c|}
\hline
  d    &$r_s=1.5$ &$r_s=2$ &$r_s=4$ &$r_s=5$  &$r_s=6$ &$r_s=7$ &$r_s=8 $  \\
\hline
0.05  & 0.21 &  --     &   --    &  --     &    --       &     --     &  --   \\
0.10  &  --     & 0.23 &   --    &  --     &    --       &     --     &  --   \\
0.15  &  --     & 0.23 &   --    &  --     &    --       &     --     &  --   \\
0.40  &  --     &  --     & 0.54 &  --     &    --       &     --     & --  \\
0.50  &  --     &  --     & 0.55 &  0.64 &    --       &     --    & --    \\
0.60  &  --     &  --     & 0.47 &  0.65 &    0.64  &    0.75 & --    \\
0.70  &  --     &   --    &  --     &  0.63 &    --      &     --     & --    \\
0.80  &  --     &   --    &  --     &  0.56 &    0.64 &    0.65  & 0.73  \\
0.90  &  --     &   --    &  --     &  0.52 &    --      &     --     & --    \\
1.00  &  --     &   --    &  --     &  0.48 &    0.57 &    0.61  & 0.63  \\
1.10  &  --     &   --    &  --     &  0.45 &    --      &     --     & --    \\
1.20  &  --     &   --    &  --     &  0.40 &    0.50 &    0.56  & --    \\
1.40  &  --     &   --    &  --     &    --    &    0,45 &    --       & 0.53  \\
1.60  &  --     &   --    &  --     &    --    &    0.41 &    0.47 & --    \\
1.80  &  --     &   --    &  --     &    --    &    --     &     --     & --    \\
2.00  &  --     &   --    &  --     &     --   &    --     &    0.40  & 0.43  \\
2.50  &  --     &   --    &  --     &     --   &    --     &    0.29  & --    \\
2.60  &  --     &   --    &  --     &     --   &    --     &     --     & 0.34  \\
3.00  &  --     &   --    &  --     &     --   &    --     &     --     & 0.29  \\
3.25  &  --     &   --    &  --     &     --   &    --     &     --     & 0.24  \\
\hline
\end{tabular}
\end{center}
\end{table}
The Monte Carlo estimate
of $h_\sigma(r)$ is $N\langle\Psi_T(R'_\sigma)/\Psi_T(R)\rangle_\pi$, where
$\Psi_T$ is the trial function, $\langle\cdot\rangle_\pi$ is an average over
$N$-particle configurations $R$ drawn
from the probability distribution $\pi$, and $R'_\sigma$ is obtained from $R$
with a random displacement ${\bf r}$ of a $\sigma$-type electron-hole pair.
Average over all  $\sigma$-type electron-hole pairs and on angles is also implemented.
We calculate both the VMC and DMC estimates $h_{\rm VAR}$ and $h_{\rm MIX}$
choosing $\pi$, respectively,
as the variational distribution $\Psi^2$ and the mixed distribution 
$\Psi_T\Phi_0$, where $\Phi_0$ is the fixed-node ground state.
Our final result is the extrapolated estimate, 
$h_{\rm EXT}=2h_{\rm MIX}-h_{\rm VAR}$, which is 
accurate to higher order in the error of the trial function than
either $h_{\rm VAR}$ or $h_{\rm MIX}$. A small correction from the
mixed to the extrapolated estimate is usually supportive of
the accuracy of both. 

An example of the computed  $h(r)$ is shown in Fig.~\ref{fig_hofr} for 
the excitonic and the quadriexcitonic phases at $r_s=5$ and $d=0.5$.
For the excitonic phase $h(r)$ is constant, within very small statistical errors,
from at least $r=30$ onwards for both VMC and DMC. This allows us to easily fit the 
tail of $h(r)$ \cite{tailfit} and infer an extrapolated estimate of the 
condensate fraction.  For the quadriexcitonic phase the pair condensate
fraction is clearly zero. We report extrapolated estimates  of the condensate fraction $n_0/n$ for $r_s\le 8$
in Tab.~\ref{table_conde}.

\section{Quadriexciton size}
\begin{figure}[]
 \includegraphics[width=.5\textwidth,trim = 10 5 0 0]{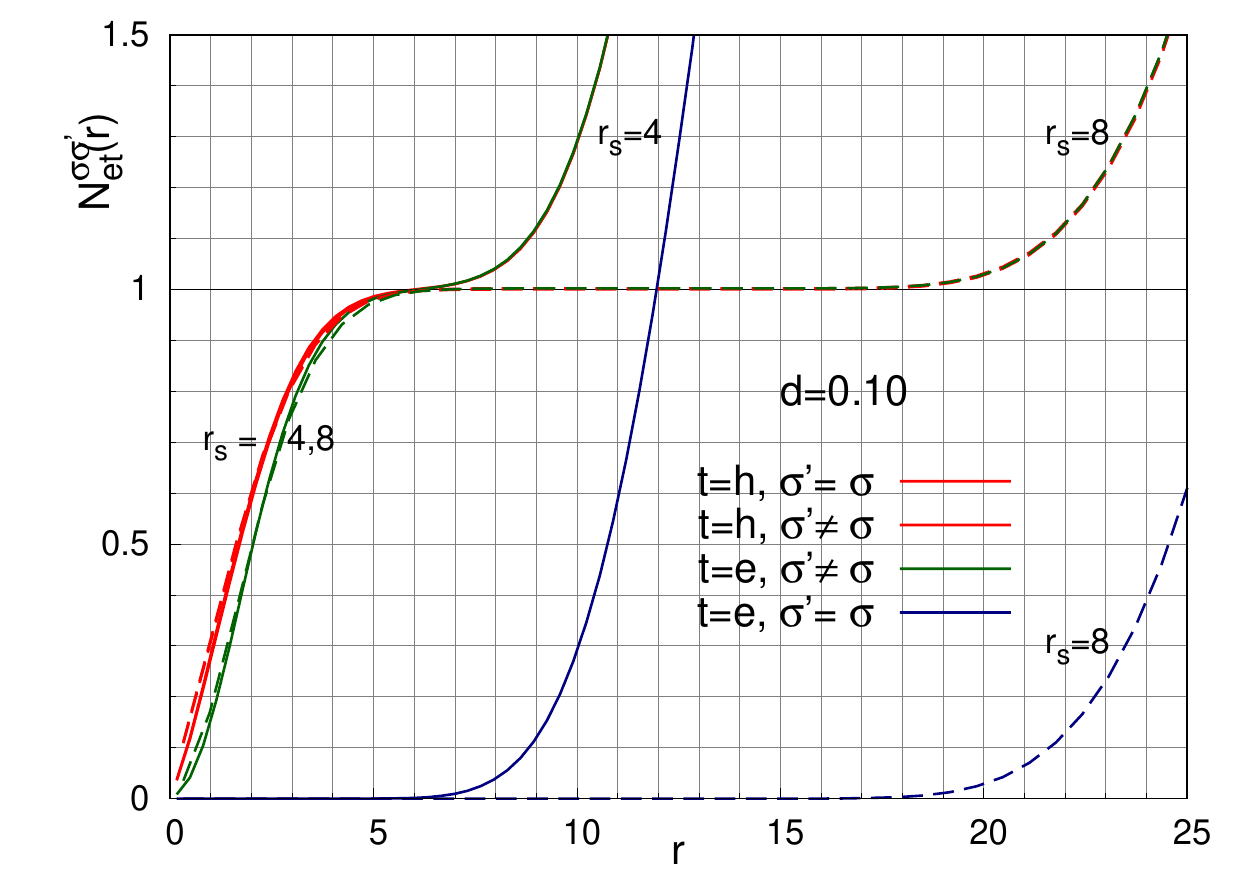}
\caption{ RCNs (extrapolated estimates), as   defined in  Eq. \ref{rcn}, at $d=0.10$ and $r_s=4$ (full 
curves), $r_s=8$ (dashed curves) }
  \label{quadri-size-fig-1}
\end{figure}
 \begin{figure}[]
 \includegraphics[width=.5\textwidth,trim = 10 5 0 0]{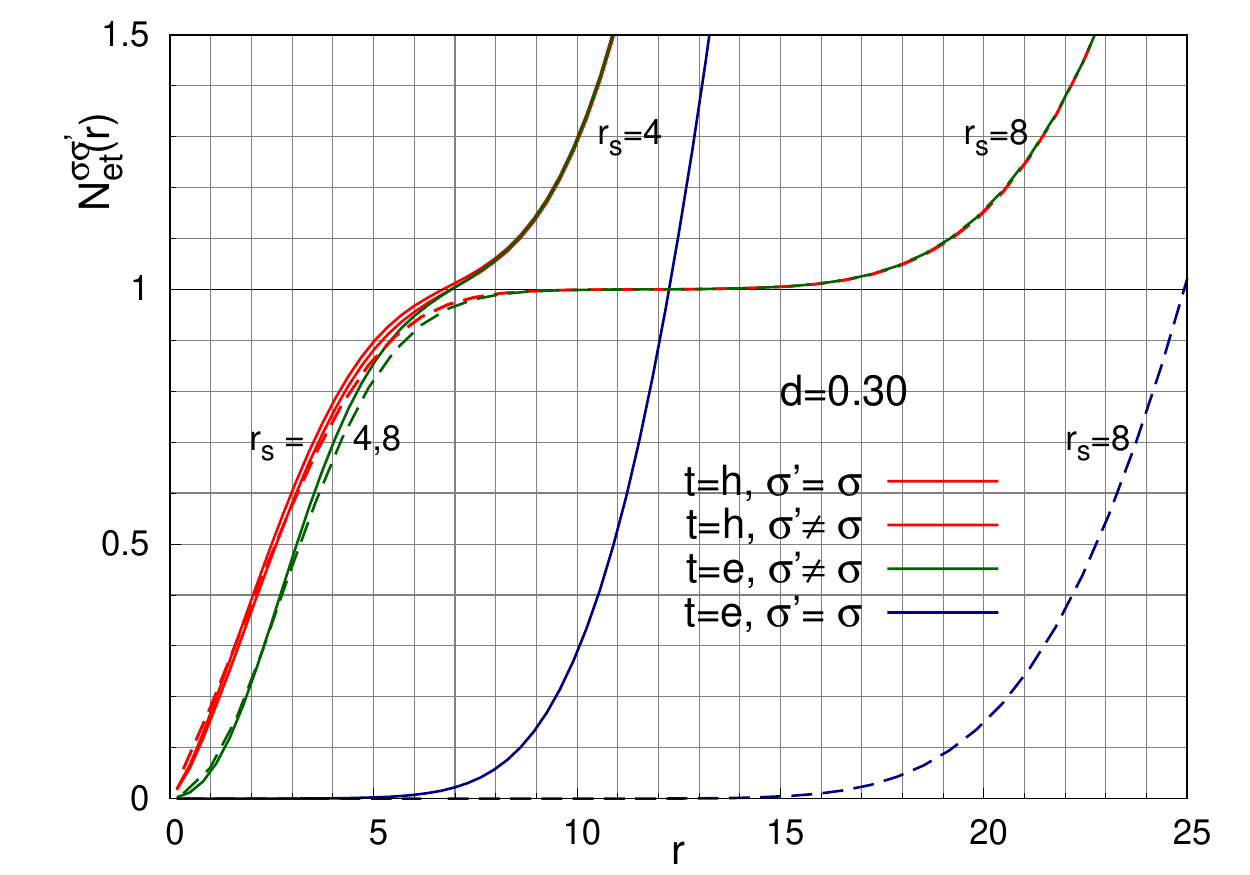}
\caption{ RCNs  (extrapolated estimates), as   defined in  Eq. \ref{rcn}, at $d=0.30$ and $r_s=4$ (full 
curves), $r_s=8$ (dashed curves) }
  \label{quadri-size-fig-2}
\end{figure}
The size of quadriexcitons and their arrangement  may be partially investigated by looking at pair 
correlation functions (PCFs) and  
the ensuing  running coordination numbers (RCNs).  The PCF  \begin{equation}
g_{tt'}^{\sigma\sigma'}(s)=\frac{1}{A}\frac{1}{n_{t,\sigma}n_{t',\sigma'}}
\left\langle\sideset{}{'}\sum_{i,j}\delta({\bf s} -{\bf r}_{t,\sigma,i}+{\bf r}_{t',\sigma',j})\right\rangle,
\end{equation}
provides the probability of finding particles of species  $(t',\sigma')$ at distance $s$ from a  particle  of  the 
species $(t,\sigma)$, with   $t=e,h$  a particle type, 
$\sigma=(v,s_z)$ a  valley-spin index, and  $n_{t\sigma}$ the areal density of the species $ (t,\sigma)$.
Above $\langle \cdots\rangle$ combines the  variational (mixed)  estimate with the rotational average. The 
extrapolated estimate for the pair correlation function is  $g_{EXT}(s)=2 g_{MIX}(s) -g_{VAR}(s)$.
 A running coordination number  (RCN) may be defined as 
\begin{equation}
N^{\sigma\sigma'}_{tt'}(r)=2\pi n_{t',\sigma'}\int_0^r\, ds\, s\, g_{tt'}^{\sigma\sigma'}(s). 
\label{rcn}
\end{equation}
Evidently  $N^{\sigma\sigma'}_{tt'}(r)$  provides the distance-dependent  pile up of particles of  of the species  
$(t',\sigma')$ around  a particle  of    the species $(t,\sigma)$ and  within a  radious $r$.  

In Fig. \ref{quadri-size-fig-1} and  \ref{quadri-size-fig-2} we display the electron-electron and electron-hole RCNs  in the quadriexcitonic phase for interplane distances $d=0.10$ and $d=0.30$, respectively, and  
densities corresponding to $r_s=4$ and $r_s=8$.  Let's focus on $d=0.10$  and $r_s=4$. It is clear that ,  
with $\sigma'\neq\sigma$,  
$N^{\sigma\sigma}_{eh}(r)$, $N^{\sigma\sigma'}_{eh}(r)$, $N^{\sigma\sigma'}_{ee}(r)$, all increasing 
functions of $r$,  cross 1 at the same point $r_q\simeq 6$ where $N^{\sigma\sigma}_{ee}(r)$ takes off from 
0. In other words, the sphere of radious $r_q=6$ contains in average exactly  4 electrons and 4 holes   (one 
for each flavour), i.e.,  a quadriexciton. We shall take $r_q$ as a gross measure of  the quadriexciton size, 
indeed an upper bound to it when $N^{\sigma\sigma}_{ee}(r)$ becomes appreciable only for $r>r_q$.  The 
next quadriexciton, as described by $N^{\sigma\sigma}_{ee}(r)$,  appears to be "touching" the one centred 
at the origin. One may resort to a crude estimate of the average distance $r_{qq}$ between nearest 
quadriexcitons as 
\begin{equation}
r_{qq} =\frac{\int_0^{r_1}ds\, s^2\, g_{ee}^{\sigma\sigma}(s)}{\int_0^{r_1} ds\, s\, g_{ee}^{\sigma\sigma}(s)}=2\pi n_{e,\sigma}\int_0^{r_1}\, ds\, s^2\, g_{ee}^{\sigma\sigma}(s),
\end{equation} 
where $r_1$ is determined by  $N^{\sigma\sigma}_{ee}(r_1)=1$. For $d=0.10$ and $r_s=4$ one obtains 
$r_{qq}\simeq 11$, which appears to be consistent with  well defined  quadriexcitons; a similar conclusion 
applies to  the case studied in the main text $d=0.05$, $r_s=4$, for which $r_q\simeq5$, while 
$r_{qq}\simeq 11$.   If one considers the case $d=0.10$, $r_s=8$ one finds $r_q\simeq 6$ and 
$r_{qq}\simeq  23$,  which is definitively consistent with well defined stable quadriexcitons.

In Fig. \ref{quadri-size-fig-2} we move to a  larger interlayer distance $d=0.30$, which  weakens  the 
quadriexciton binding. Indeed  at $r_s=4$ we observe the relaxation to a larger size  $r_q=7$, as compared 
with $r_q=6$ found for   the  $d=0.10$ case,   a minor mixing between neighboring quadriexcitons,  and still 
$r_{qq}\simeq 11$.  Lowering the density to   $r_s=8$ we find an appreciable increase in the quadriexciton 
size, from $r_q=6$  at $d=0.1$ to $r_q=9$, as well as an increase in the distance between quadriexcitons 
with  $r_{qq}\simeq 29$. However at $r_s=8$ quadriexcitons appears again well separated and thus well 
defined.

Thus, it would appear that  the  size $r_q$   of quadriexcitons  (i)  increases with $d$, at given $r_s$, due to 
the weakening of the binding, and (ii) it  may increase with lowering the density (increasing $r_s$), at given 
$d$, if one starts from a situation  where there is some mixing between neighboring quadriexcitons.   
Interestingly, if we extrapolate to $d=0$ the quadriexciton  size calculated for finite $d$ at  $r_s=4$, we find 
a value comparable to the biexciton size $r_q\simeq 4$ found at $d=0$   in the one valley system 
~\cite{maezono2013}.